\documentclass[sigconf]{acmart}
\usepackage{booktabs} 

\renewcommand\footnotetextcopyrightpermission[1]{} 
\setcopyright{none} 

\usepackage[utf8]{inputenc}
\usepackage[USenglish]{babel}
\makeatletter

\@namedef{ver@fixltx2e.sty}{2006/03/24}
\@namedef{opt@fixltx2e.sty}{}

\@namedef{ver@multicol.sty}{2016/04/07}
\@namedef{opt@multicol.sty}{}

\PassOptionsToPackage{hyperfootnotes=false}{hyperref}

\def\@makefnmark{\hbox{\@textsuperscript{\normalfont\@thefnmark}}}

\makeatother

\usepackage{algorithm} 
\usepackage{algpseudocode}
\usepackage{xspace}
\usepackage{subcaption}
\usepackage{amsmath,amsfonts}
\usepackage{multirow}

\def\etal{\emph{et al.}}
\def\ie{\emph{i.e.}}
\def\eg{\emph{e.g.}}

\usepackage{color}
\definecolor{Orange}{rgb}{0.9,0.5,0}
\definecolor{NavyBlue}{rgb}{0.1, 0.4, 0.8}
\definecolor{Magenta}{rgb}{0.8, 0.1, 0.6}
\definecolor{Green}{rgb}{0.1, 0.8, 0.3}
\definecolor{DarkGreen}{rgb}{0.0, 0.7, 0.2}
\definecolor{Brown}{rgb}{0.4, 0.3, 0.1}
\definecolor{Burgundy}{rgb}{0.5, 0.0, 0.13}
\definecolor{BrightCerulean}{rgb}{0.11, 0.67, 0.84}
\definecolor{BlueViolet}{rgb}{0.33,0.1,0.5}

\newcommand{\paragraphbe}[1]{\vspace{0.75ex}\noindent{\bf \em #1}\hspace*{.3em}}

\makeatletter
\def\plist@algorithm{Alg.\space}
\makeatother

\newcommand{\system}{{PoliFL}\xspace}
\newcommand{\systemcentral}{{\system Server}\xspace}
\newcommand{\databox}{{Databox}\xspace}

\newcommand{\folder}{./figures}

\begin{document}



\author{Kleomenis Katevas}
\affiliation{Telefonica Research}
\email{kleomenis.katevas@telefonica.com}

\author{Eugene Bagdasaryan}
\affiliation{Cornell Tech}
\email{eugene@cs.cornell.edu}

\author{Jason Waterman}
\affiliation{Vassar College}
\email{jawaterman@vassar.edu}

\author{Mohamad Mounir Safadieh}
\affiliation{Vassar College}
\email{msafadieh@vassar.edu}

\author{Eleanor Birrell}
\affiliation{Pomona College}
\email{eleanor.birrell@pomona.edu}

\author{Hamed Haddadi}
\affiliation{Imperial College London}
\email{h.haddadi@imperial.ac.uk}

\author{Deborah Estrin}
\affiliation{Cornell Tech}
\email{destrin@cs.cornell.edu}

\title{Policy-Based Federated Learning}

\begin{abstract}
{

In this paper 
we present \emph{\system}, a decentralized, edge-based
framework that supports heterogeneous privacy policies for federated 
learning. We evaluate our system on three use cases that train models
with sensitive user data collected by mobile phones---predictive text, 
image classification, and notification engagement prediction---on a Raspberry~Pi edge device. We
find that \system is able to perform accurate model training and 
inference within reasonable resource and time budgets while also 
enforcing heterogeneous privacy policies.}
\end{abstract}


\maketitle
\pagestyle{plain}

\section{Introduction}
\label{sec:introduction}

We are surrounded by an increasing variety of smartphone apps and Internet of Things (IoT) devices, which pervasively collect a wealth of personal data. 
These data can be used to develop machine learning models that support a wide variety of features,
including content personalization~\cite{contentpers}, health analytics~\cite{health_analytics},
and image classification~\cite{image_class}. However, much of this data is highly sensitive---it
can be used to infer individual behavioral patterns, health, preferences, activities,
and personal relationships---so leveraging the full potential of this data
depends on developing tools that guarantee the privacy of sensitive data.

Most current machine learning systems are designed to centrally collect, aggregate,
and process data, typically on some kind of cloud-hosted service; 
this centralized approach fails to guarantee adequate privacy protection
because users must trust the centralized service provider to not misuse data
and to secure data against unauthorized 
disclosure~\cite{breaches}. 
Federated machine learning~\cite{mcmahan2017communication,kairouz2019advances}---an example of \emph{edge computing} in which data is processed (\ie,~the model is trained) as 
close to the data source as possible---offers a promising alternative for 
designing privacy-enhancing machine learning systems. Sensitive 
data are processed locally on a user's phone~\cite{hard2018federated,leroy2019federated}
or IoT device~\cite{nguyen2019diot,mills2019communication}, thereby 
bypassing many of the security and privacy challenges posed by such a 
service. As is the case for machine learning, sensitive data that are used to train a federated learning model can be leaked through model
inversion attacks~\cite{fredrikson2014privacy,fredrikson2015model}, but this
threat can be mitigated through the use of differential privacy~\cite{dwork2008differential,geyer2017differentially,wei2020federated}.
Recently, large service providers like Google and Apple have adopted federated learning as a solution for deploying scalable algorithms for on-device analytics within highly sensitive tasks~\cite{bonawitz2017practical,47976, kairouz2019advances}.

\begin{figure*}[t]
  \centering
  \includegraphics[width=\linewidth]{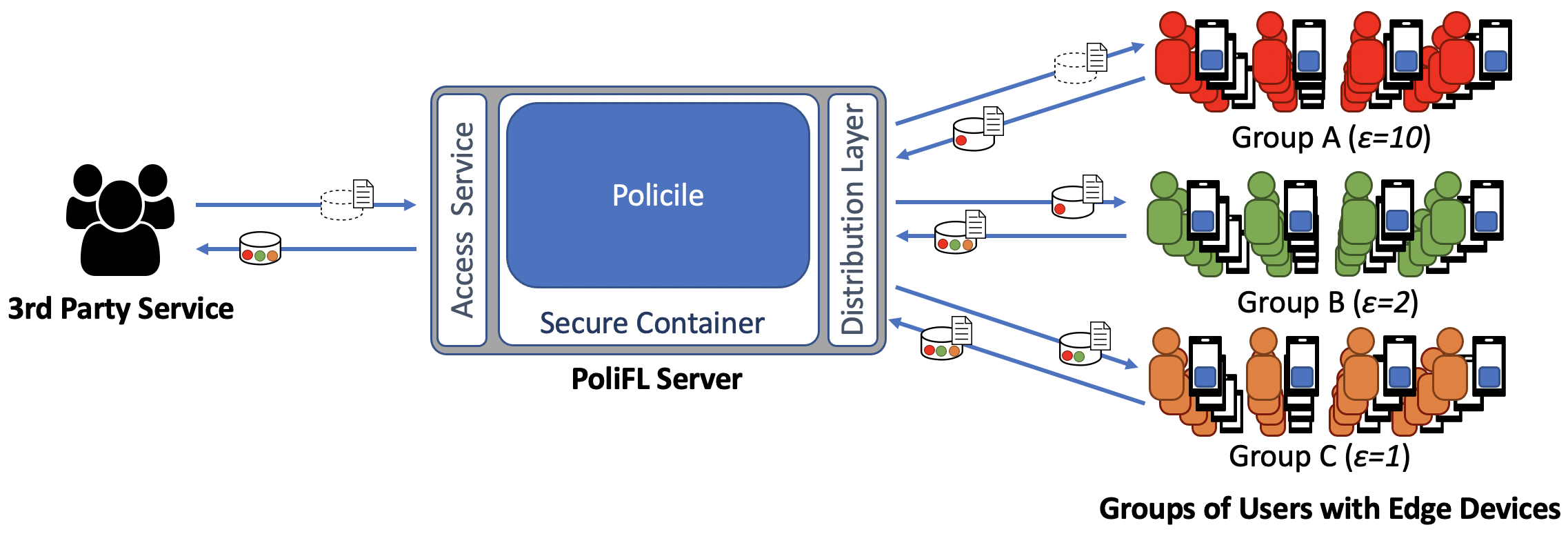}
  \caption{\textbf{The PoliFL platform}. A third-party service sends a request (serviced by the ``Access Service'' web-module) for a generalized ML model trained with multiple users. A policy that conforms with the request is securely distributed (using an encrypted VPN provided by the \emph{Distribution Layer}) among the device groups following the concept of cascaded training. In each round, the policy is executed (\ie,~model is trained using specific data sources as enforced by the policy) within the device's instance of Policile (blue box), and the resulting models are returned to the \systemcentral where the server's instance of Policile applies federated averaging and protects the merged model with differential privacy of the group's $\epsilon$ budget. The same process is repeating until all groups are consumed. The final model is returned to the interested third-party service.}
  \label{fig:overview}
\end{figure*}

However, prior work on federated machine learning has assumed that data
collection and use are governed by a single privacy policy on homogeneous data (\eg,~models are 
trained either entirely with or entirely without
differential privacy, and with uniform privacy parameters), an assumption
that is inconsistent with the federated model. In practice, even a centrally-run
federated learning system would need to support users with different privacy requirements either because the system exposes some privacy settings to the user---\eg,~allowing users to opt-in for ``high'' privacy guarantees or continue with the ``default'' privacy settings, or allowing users to allow or disallow the use of particular data
sources such as location data or microphone data---or because different regulations (\eg,~HIPAA~\cite{annas2003hipaa} or GDPR~\cite{gdpr}) impose different privacy requirements for data about users from different geographic jurisdictions.
In a \emph{federated} federated
learning system, different organizations within the federation (\eg,~different apps) would  likely 
offer their users slightly different privacy guarantees, all of which must by followed by any federated learning application that uses their data. 


In this work, we develop \emph{\system}\footnote{The complete source code of the \system system, evaluation scripts and data are available at \url{https://github.com/minoskt/PoliFL}.} to support heterogeneous privacy policies for federated learning (Fig.~\ref{fig:overview}). Building on prior 
work in edge computing~\cite{10.7146/aahcc.v1i1.21312} and use-based
privacy~\cite{bagdasaryan2019ancile}, \system offers a scalable, 
privacy-enhancing approach to model training data processing that enforces non-uniform privacy requirements. In \system, data are processed locally
at the edge, and each local data source enforces 
user-specific data use policies as data are processed (\eg, during the 
local model training phase). A central service called the \emph{\systemcentral} is responsible for 
receiving data processing requests from third-party services and 
coordinates data access at each edge node; data are then accessed and processed locally 
in accordance with the user's local policy. If authorized by the policy,
the processed results (\eg,~a locally
trained model) are sent back to the central service, which aggregates 
the results from all users and then returns the final result to the 
requester. The policy language enforced by \system is defined in Sec.~\ref{sec:policy-federated-learning} and the design and implementation of the system are described in Sec.~\ref{sec:platform}.

To demonstrate the utility of our system and evaluate its performance, we identify three different use cases in which machine learning models are trained on sensitive data collected by mobile phones:

\begin{enumerate}
	\item \textbf{Language Modeling.} Based on text that a user has typed so far, this model predicts the next word that the user would like to write. This sort of model is used to facilitate text entry in email and messaging apps on mobile phones.
	
	\item \textbf{Image Classification.} Given a photograph, this model classifies the object visible in the photo. Such models can be used to automatically generate photo albums, and is a benchmark machine learning task that has been used in other federated learning works~\cite{zhao2018federated,bagdasaryan2020backdoor,wang2019adaptive}. 
	
	\item \textbf{User Behavior Modeling.} Given a comprehensive 
	suite of mobile sensor data (including data from sensitive sources such as location, microphone, and call records),
	this model predicts whether a user will interact with a notification at a 
	particular point in time. This sort of model could be used to minimize 
	disruptions by and streamline the efficiency of mobile phone notifications.
\end{enumerate}

In Sec.~\ref{sec:system_performance}, we evaluate the performance of \system to determine the (1) internal overhead of the edge components, (2) the external overhead imposed by the network traffic, and (3) the scalability of our system. We find that policy-compliant federated learning is feasible for limited-resource edge devices such as the Raspberry Pi, and we find that the external overhead is reasonable relative to the time to compute the model. Leveraging our three example use cases, we also find that our system scales to support training with 100 edge devices per FL round; the majority of the overhead was incurred by the time to compute the model on the edge devices.

In Sec.~\ref{sec:use-cases} we explore the effect of heterogeneous policy-compliance on the accuracy of the resulting federated learning models for each of our three use cases. We consider three policy compliance strategies: training on a subset of the data with a single uniform policy, training on the full dataset while enforcing the join of all applicable policies, and training using a new technique we introduce called \emph{cascaded policy compliance}. In all cases, we are able to achieve accuracy rates of 14-17\%; in two of our use cases, higher accuracy was achieved through cascaded training than is achieved by the baseline strategies. These results suggest that supporting heterogeneous policy enforcement might be an effective approach to maximizing the performance of privacy-preserving machine learning applications.

We conclude that \system, the first system for enforcing non-uniform privacy policies for federated learning, constitutes a key step towards supporting high-performance, privacy-preserving machine learning applications.

\section{Background and Related Work}
\label{sec:related_work}

\subsection{Privacy at the Edge} There has been a large body of research demonstrating how \emph{edge computing} and local analytics can help in preserving data privacy while enabling accurate analytics~\cite{10.1145/2831347.2831354, 47976, 10.7146/aahcc.v1i1.21312}. A number of recent industry efforts have also adopted this model for performing accurate analytics while respecting privacy. For example, the Brave browser delivers personalized adverts based on analytics performed locally on users' devices~\cite{toubiana2010adnostic, guhaserving}. The BBC has also developed the BBC~Box\footnote{\url{https://www.bbc.co.uk/rd/blog/2019-06-bbc-box-personal-data-privacy}} for private and secure aggregation of data and content personalization.




\subsection{Privacy Preserving Machine Learning} 

Data used to train machine learning models, such as the data passively generated by smartphones, can reveal sensitive details such as behavioral
patterns~\cite{washpost, handel2013smartphone} and physical presence
\cite{wicker2012loss}; the release of such details can lead to stalking or disparate
treatment~\cite{woodlock2017abuse,handel2013smartphone}.
There have been several efforts to support improved privacy guarantees for smartphones and other ubiquitous devices \cite{li2017privacystreams,enck2014taintdroid,arzt2014flowdroid,wang2015using,pournaras2015privacy, li2018coconut,
narain2019mitigating}, however these efforts do not enforce general data use restrictions and most do not support machine learning applications. 


One of the first
techniques that was used for protecting the user's privacy while running
joint computation is Secure Multi-Party Computation
(MPC)~\cite{goldreich1998secure}, a cryptographic technique that allows
mutually distrusting parties to run joint computations without revealing
private data. One of the main limitations of such technique is that it
performs badly with large datasets, even with only a few parties getting
involved. Volgushev~\etal~\cite{volgushev2019conclave} recently
presented a scalable solution, based on MPC, that is applicable to large
datasets, using query rewriting to minimize expensive processing under
MPC. Other approaches such as Opaque~\cite{zheng2017opaque} and DarkneTZ~\cite{mo2020darknetz} use a Trust Execution
Environment (TEE) to run most of the computation under a secure
environment.

Other works have focused specifically on applications of personal model training~\cite{servia2018privacy}, either in the general context of privacy-preserving ML for training on encrypted data~\cite{PPMLaaS}, or specific to the context of Federated Learning (FL) where independent third-party apps collaborate and combine their data and models locally for building joint meta-models in an FL fashion~\cite{flaas}. 

\subsection{Federated Learning}

\emph{Federated Learning (FL)} is a novel technique to perform distributed training at the edge~\cite{47976,kairouz2019advances,fedlearn_1, fedlearn_2}; models are trained in a decentralized fashion without the need to collect and process the user data centrally. The global provider distributes a shared ML model to multiple users for training on local data, and then aggregates the resulting models into a single, more powerful model, using \emph{Federated Averaging}~\cite{fedlearn_1}.

\begin{table}[t]
\centering
\caption{Federated learning definitions.}
\label{tab:fl_definitions}
    \begin{tabular}{ll}
    Variable & Description \\
    \midrule
    $G^t$ & global model at round $t$. \\
    $n$ &  total number of participants. \\
    $m$ &  subset of participants selected for a single round. \\
    $\eta$ & global learning rate. \\
    \textit{L} & locally trained model. \\
    $\mathcal{D}$ & local data. \\
    \textit{E} & number of epochs for local training. \\
    \textit{lr} & local learning rate.\\
     $S$ & clipping bound. \\
     $\sigma$ & amount of added noise. \\
\end{tabular} 
\end{table}

More particularly, it randomly selects a subset of $m$ participants from the total participants $n$ and sends them the current joint model $G^t$ in each round $t$. Choosing $m$ involves a trade-off between the efficiency and the speed of training. Each selected participant updates this model to a new local model $L^{t+1}$ by training on their private data and sends the difference $L_i^{t+1} - G^t$ back (see notation in Table~\ref{tab:fl_definitions}). Communication overhead can be reduced by applying a random mask to the model weights~\cite{fedlearn_2}. The central server averages the received updates (or uses other aggregation methods~\cite{reddi2020adaptive}) to create the new joint model:

\begin{equation}
\label{eq:1}
G^{t+1} = G^{t} + \frac{\eta}{n}\sum_{i=1}^m (L_i^{t+1} - G^t)
\end{equation}


To further
protect user data, recent solutions extend federated learning with
differential privacy~\cite{fedlearn_dp} or
encryption~\cite{bonawitz2017practical, hardy2017private}. These
approaches are relevant to our work, however the privacy policies cannot
be dynamically negotiated and are uniform across users. Instead, we
investigate a policy-controlled execution of federated learning, where
each user has a personal policy that specifies how to handle their data
depending on the application specifics.

\subsection{Differential Private Federated Learning}

While the provider never accesses the user data, previous works have shown that the parameters of the model can leak sensitive information about the user~\cite{hitaj2017deep, li2013membership, melis2019exploiting, salem2018ml}. This can lead to significant privacy threats such as membership inference attacks where an attacker can determine if a particular user was part of the model's training set~\cite{shokri2017membership, yeom2018privacy}. Differential Privacy (DP)~\cite{dwork2008differential} has been proposed by researchers as a privacy guarantee from such type of attacks.

Differential privacy~\cite{dwork2008differential, dwork2011differential} is a common technique to preserve privacy of a single entry in the dataset on some query.  We adopt $(\epsilon,
\delta)$-differential privacy which is already used in 
deep learning applications~\cite{abadi2016deep}. A randomized mechanism
$\mathcal{M}: \mathcal{D} \rightarrow \mathcal{R}$ with a domain
$\mathcal{D}$ and range $\mathcal{R}$ satisfies $(\epsilon,
\delta)$-differential privacy if for any two adjacent datasets $d,d'
\in \mathcal{D}$ and for any subset of outputs $S \subseteq \mathcal{R}$:
$$
    Pr[\mathcal{M}(d) \in S] \leq e^{\epsilon} 
    Pr[\mathcal{M}(d') \in S] + \delta   
$$
When implementing differential privacy, it is necessary to set a
\emph{privacy budget} and every computation charges an
$\epsilon$ cost to this budget; once the budget is exhausted, no further
computations are permitted on this dataset. 

In machine learning, the mechanism $\mathcal{M}$ represents a training procedure and the result $S$ represents a model. Abadi~\etal~\cite{abadi2016deep} propose moments accountant and differentially private stochastic gradient descent (DPSGD) to perform training and compute the total cost of producing a model. DPSGD clips the model gradients to norm $S$ and adds Gaussian noise with standard deviation $\sigma$ to obtain a differentially private model.

Recent work~\cite{fedlearn_dp} presented differentially private
federated learning with a participant level privacy. Similar to DPSGD,
each participant's update is \emph{clipped} to the norm $S$, \eg,~
multiplied the value by $\textnormal{max} (1, \frac{S}{||L^{t+1}_i - G^t||_2})$,
to bound the sensitivity of the updates. Additionally, Gaussian noise
$\mathcal{N}(0,\sigma)$ is added to the weighted average of updates:

$$G^{t+1} = G^{t} + \frac{\eta}{n}\sum^m_{i=1} (\texttt{Clip}(L^{t+1}_i-G^t, S) + \mathcal{N}(0,\sigma))$$.

This design assumes clipping models locally and then adding noise centrally. 
In a simplified design that users are capable of adding noise locally $\mathcal{N}(0, \sigma^2)/m$ (see Algorithm~\ref{alg:traindp}) such that accumulated noise in a round sums up to $\mathcal{N}(0, \sigma^2)$. An alternative design, implemented in TensorFlow Federated~\cite{tff}, only performs local clipping but modifies the accumulate function to add sufficient noise to guarantee differential privacy. Both designs are supported by our implementation.

\subsection{Enforcing Policy-based Privacy.} Language-level information flow
control techniques~\cite{sabelfeld2003language} is a popular approach on
tracking data propagation through the system. The recently proposed
Ancile framework~\cite{bagdasaryan2019ancile} combines this concept with
use-based privacy~\cite{birrell2017reactive} that introduces policies
that ``react'' to data transformation. However, that framework is
limited to centralized data processing and does not support distributed applications such as federated learning. 

Recent work~\cite{fischer2020computation}, proposed data flow
authentication control using homomorphic encryption and SGX enclaves
that prevents an adversary from modifying the program. However, this
approach requires a fixed program to be authorized, whereas our
framework supports flexible programs that satisfy the corresponding
policies. 

This work extends these prior works to provide flexible, non-uniform policy enforcement for federated learning.

\section{Policy-Based Federated Learning}
\label{sec:policy-federated-learning}

In this section, we describe our trust model, define our policy language, give examples of privacy policies for federated learning, and explore how policy-compliance affects performance for three different federated learning use cases. 

\subsection{Trust Model}

There are three classes of principals in a typical federated learning system: 
\begin{itemize}
    \item \textbf{Users:} generate sensitive but useful data and have
    certain expectations for data usage that could be inherited or assigned by an authority.
    \item \textbf{Applications:} provide a service that requires a
    model to be trained on user data.
    \item \textbf{FL Platforms:} maintain the infrastructure to train
    federated models that are deployed by a trusted entity. These are distributed platforms
    that include both a \emph{coordination} server and code that runs on a device. 
\end{itemize}

We assume that users trust the FL Platform to monitor the
application's usage of personal data and enforce policy compliance. 
Note that in systems 
where the application and the FL Platform are produced by the same company (as is common 
today), this trust model still improves data privacy by reducing the trusted computing 
base and separating policy compliance from application code. 
While it is theoretically possible to eliminate trust in the central component of the FL platform by implementing local differential privacy on the edge devices, we do not implement this approach due to performance considerations~\cite{naseri2020toward}.

\paragraphbe{Heterogeneous policies.} Unlike prior systems, we do not assume that all 
users have a single policy governing how their data may be used. Instead, we assume that each 
user adopts one of a small set of available policies either by modifying their preferences
(\eg,~opting in or out of use of particular data sources such as location or microphone, or use a stronger privacy guarantee) or by
selecting a service provider (\ie,~application) within a federated system. 

\paragraphbe{Limitations of the trust model. }
We assume that principals
might need full access to the aggregated model to test it
and distribute it to new users who did not participate in the training. We therefore do not consider fully
encrypted training like SecureML~\cite{mohassel2017secureml}, which could also eliminate the need to trust the FL platform. 
However, we observe that techniques such as secure
aggregation~\cite{bonawitz2017practical} that decrypt the aggregated model for the FL Platform can be deployed.

\subsection{A Policy Language for FL}

Federated learning limits the attack surface by exposing raw user data only to the application running on the device; privacy guarantees therefore depend on preventing data misuse by the local application.

Restrictions that focus on preventing unauthorized data use can be best expressed as \emph{use-based privacy polices}. In use-based privacy, data are associated with policies that authorize certain types of data use while denying other uses. These policies are reactive, as they can change as data is transformed. For example, a use-based policy for location data might allow the current smartphone location to be viewed by a weather application only at the granularity of a city; this policy would specify that raw location data may be used to determine which city the phone is currently in, and that the city-level location may be used by the weather app in any way. Policies may also change due to external events. For instance, a use-based location policy might state that a user location may only be shared while the user is driving a company vehicle. The ``use'', whether the location data can be shared, changes based on the event of driving the company car. 

In this work, we express privacy policies as use-based privacy policies using the policy language developed for Ancile~\cite{bagdasaryan2019ancile}. Policies are specified as regular expressions over an ``alphabet'' of
commands.  Policies define a state transition diagram, specifying how
data values and any derived values may be used.  For example, if the command \texttt{fetch\_data} has to
be followed by the \texttt{filter()} command before it can be publicly
released, the policy will look like:
$$\texttt{\textbf{fetch\_data}}~.~\texttt{\textbf{filter(col=`location')}}~.~\texttt{\textbf{return}}$$
As shown above, our policy language supports function parameters, \eg,~filtering a location column for the data, and also supports logical operations, \eg,~sequential composition (\texttt{.}), union (\texttt{+}), iteration (\texttt{*}), and negation (\texttt{!}).  Once the execution reaches $\texttt{return}$, a program is authorized to use data without any further restrictions. Appendix~\ref{appx:policyenforce} provides more details about the policy language.


\subsection{Example Policies for FL}
\label{sec:policy-federated-learning:ubp_fl}

In our system, we assume that not all users will necessarily have the same privacy policy. Non-uniform policies might arise if some users opt-in to stronger privacy protections while others accept the application default, if model training combines data from multiple federated platforms (\eg,~different hospitals or different phone providers) with different policies, or if users are subject to policies imposed by different legal jurisdictions. 

In this section, we describe three classes of policies that might apply to a subset of the users of a system: (1) a policy that authorizes federated learning, (2) a policy that authorizes federated learning with differential privacy (for some $\epsilon$ budget), and (3) a policy that restricts which information can be used to train a federated learning model. 

\paragraphbe{Authorizing Federated Learning.} A simple, high-level policy that authorizes data to be used to train a model in a federated learning fashion might be: 

\begin{align*}
&\texttt{{\color{blue}get\_data}~.~{\color{red}runFL}~.~return}
\end{align*}

A function \texttt{get\_data()} is executed locally on each of the user devices, where a high-level function \texttt{runFL()} is executed on both the server and the user devices before the resulting model can be returned to the application. We color code local execution as blue and server execution as red.

Alternatively, a privacy policy can be specified with more granularity. Federated learning applications leverage three key steps: local training, accumulation of local models, and averaging or aggregation of 
the accumulated sum (see Algorithms~\ref{alg:trainlocal}-\ref{alg:average}). A privacy policy can therefore be specified in more detail by expanding
\texttt{runFL()} to:

\begin{align*}
& \texttt{{\color{blue}get\_data} . {\color{blue}train\_local}. 
{\color{red}accumulate$^*$}}\;.\\
& (\texttt{{\color{blue}train\_local}\;.\; {\color{red}accumulate$^*$} + {\color{red} average$^*$})\;.\;return}
\end{align*}

This policy, first, ensures that user's data is only used in local
training on the device. Second, it specifies that the result of local
training can be combined with other models and the resulting model can
participate in future iterations of the federated learning, before being
returned to the third-party provider. 
By using iteration in the policy, we support multiple rounds of training
using the combination of policies. Function
\texttt{\texttt{train\_local}} takes both the current model $G^t$ with
policy $P_{model}$ and local user data $\mathcal{D}$ with policy
$P_{data}$ and produces a new model $L^{t+1}$ with the policy $P_{model}~\&~P_{data}$. Similarly, functions \texttt{\texttt{accumulate}} and
\texttt{\texttt{average}} combine two policy-controlled objects. We
prevent explosion of policy size by using simple reduction rules that
keep the policy size constant~\cite{bagdasaryan2019ancile}. 

\paragraphbe{Differential privacy for FL.} If the user belongs to a group that has tighter privacy restrictions, their policy might require the resulting global model to be differentially private, \eg,~enforce a check on the spent budget:

\begin{align*}
& \texttt{{\color{blue}get\_data}\;.\; {\color{red}runFL}}\;.\; \\
& \texttt{{\color{red}\texttt{enforce\_dp\_budget(eps=1)}}\;.\;return}
\end{align*}

This check happens on the server side since the differentially private guarantees are central and with our current threat model, they are only important when the model is shared with the application. In this setting the FL platform still observes the raw user updates which might still pose a privacy concern. In order to also prevent the FL platform from learning user data, it is possible to use local differential privacy as part of a policy. However, current performance of models trained with local DP is extremely low~\cite{naseri2020toward}. A program that complies with the DP policy can use the \texttt{local\_dp\_train()} function from Algorithm~\ref{alg:traindp}:

\begin{align*}
& \texttt{{\color{blue}get\_data} . {\color{blue}train\_local\_dp}. 
{\color{red}accumulate$^*$}}\;.\\
& (\texttt{{\color{blue}train\_local\_dp}\;.\; {\color{red}accumulate$^*$} + {\color{red} average$^*$})\;.}\\
&  \texttt{{\color{red}\texttt{enforce\_dp\_budget(eps=1)}}\;.\;return}
\end{align*}

\paragraphbe{Access control for FL.} Another common example is to restrict
certain data fields to be accessed by the program. A user might
not allow access to specific data sources (\eg,~location and microphone). A policy attached to user data will be required to check the conditions and perform local data filtering before proceeding with model training:
\begin{align*}
& \texttt{\color{blue}get\_data\;.\;filter(sensors=[`mic', `loc'])}\;.\; \\
& \texttt{{\color{red}\texttt{runFL}}\;.\;return}
\end{align*}

Unlike simple 
field protection, use-based privacy allows one to specify the context and conditions~\cite{bagdasaryan2019ancile} under which the data can be available. For example, to only collect data when a user is at the
specific location, \eg,~on campus, at work, etc:

\begin{align*}
& \texttt{\color{blue}get\_data\;.\;in\_geofence\_cond(geofence=GF)\;.} \\
& \texttt{{\color{red}\texttt{runFL}}\;.\;return}
\end{align*}

\section{The \system Platform}
\label{sec:platform}



In this section we describe the details of the \system platform, a decentralized policy-based system for running FL tasks. We assume that a third-party service is interested in building a generalized ML model from a collective policy-enforced personal data processing. We also assume a heterogeneous preference on the level of privacy, such as where users are classified into several privacy groups based on the $\epsilon$ preference budget or based on sensor data access. Fig.~\ref{fig:overview} gives an overview of \system and its components. The \system Platform consists of a \systemcentral and multiple edge devices.  The \systemcentral is responsible for coordination with the edge devices, which compute local models on sensitive user data.  Policy enforcement is performed by an instance of the Policile policy enforcement engine, running on both the \systemcentral and the edge devices, and is described below.

\subsection{Policile framework}
\label{sec:platform:Policile}

Policy enforcement is provided by Policile, an extension of the Ancile platform~\cite{bagdasaryan2019ancile} to support distributed, policy-based FL.  To train an ML model, a third-party service submits a program to the \systemcentral, which runs the program in the Policile trusted environment.

The core programming abstraction of Policile is a distributed \emph{Data-Policy Pair}, an opaque object containing data (such as an FL model or device data used to build a model) and an associated policy written in the language described in Sec.~\ref{sec:policy-federated-learning:ubp_fl}. 

To enforce policy compliance, submitted programs are compiled using the RestrictedPython~\cite{restrictedpython} language-level framework before they are executed.  RestrictedPython allows access to Data-Policy Pairs only through trusted library functions which are checked for policy compliance when they are called.  Therefore, it is not possible to access user data outside of what is authorized by the policy associated with the Data-Policy Pair.  Our design allows the FL platform to avoid manual inspection of the third-party's code by relying on the policies attached to users' data instead.

Our work extends the enforcement of these policies across the whole FL platform, spanning both edge devices and the coordinator server (\ie,~\systemcentral in the context of our work). Data-Policy Pairs are now distributed and as such, can be transferred to and from the \systemcentral and the edge devices. In order to prevent inspection or manipulation of Data-Policy Pairs outside of authorized library functions, Data-Policy Pairs are transferred between \systemcentral and the edge devices internally through a WireGuard\footnote{https://www.wireguard.com} Virtual Private Network (VPN).  Furthermore, instances of Policile on both the \systemcentral and the edge devices are run from inside of Docker containers to ensure data are always protected by policies that enforce the type of access across the system.


\subsection{\systemcentral}
\label{sec:platform:PoliFLServer}

The \systemcentral is responsible for maintaining data policies and coordinating the communication between a third-party service and all interested devices. It runs as a web-service and receives programs on behalf of the third-party services. These services communicate with the \systemcentral by making model requests through the \emph{Access Service} module shown in Fig.~\ref{fig:overview}.

The \systemcentral maintains data policies for all users registered in the system. A web dashboard allows the \systemcentral administrator to view, add, delete, and modify the policies that are available to each of the interested third-party services. When a third-party service requests a generalized ML model from the registered edge devices, the service sends a request with the following elements:

\begin{itemize}

\item {\bf Application Token:} A secret token that is used to
authenticate the service to the \systemcentral.

\item {\bf Global Program:} This program aggregates the results (locally trained models) from the
edge devices and applies differential privacy to the federated averaged model. The result of this program, if policy compliant, will either be distributed again to the next group of edge devices, or be returned to the service who made the request.

\item {\bf Local Program:} A piece of computation to be executed on the
edge devices and whose result, if policy compliant, will be returned to
the \systemcentral.

\end{itemize}

The \systemcentral handles each request by first validating the
service's token. Once the service has been validated,  Policile executes
the Global Program. The Global Program then selects the participants and
sends the Data-Policy Pair, described in Sec.~\ref{sec:platform:Policile}, along with the Local Program to the selected
users on the edge devices. Once the edge device receives the request from the \systemcentral, the Local Program is
executed on the Policile instance of the edge device. 


\subsection{Edge Device}
\label{sec:platform:edgedevices}

We assume that an edge device exists per \emph{user} in the form of a physical device (\eg,~a mobile phone or a home gateway) that stores data related to the interested FL task. Each device has a secure container that provides isolation from the operating system environment (including controlled network access, system resources etc.). It also includes a deployment of Policile which ensures that all edge computation and data access is done in a policy compliant manner.

As a secure data container, our current implementation utilizes \databox~\cite{10.7146/aahcc.v1i1.21312}, a Docker-based~\cite{containers} secure open-source platform that enables data subjects to manage controlled access by third parties to their personal data. For code in a container to access personal data requires relaxing a
``default deny'' configuration, done on the basis of user-authorized
permissions represented as bearer tokens granted by the system to the
container in question. Each \databox container comes with a manifest
that outlines the permissions it requires, which is translated on
installation into a \emph{service level agreement} (SLA) encoding the
granted data access and export permissions, which the platform then
enforces.

Note that, when allowed by the policy, the edge device is allowed to send Data-Policy Pairs (\eg,~a locally trained ML model) back to the \systemcentral, where then can be used by the \systemcentral Global Program (\eg,~for federated averaging).

\section{System Performance}
\label{sec:system_performance}

In this section, we seek to answer the following performance question about our system:
\begin{itemize}
  \item What are the system overheads of the \systemcentral and edge components?
  \item What are the networking overheads imposed by the system?
  \item What is the performance of our system training  multiple simultaneous users for a round of training? 
\end{itemize}

In order to answer these questions we conducted evaluations using
three use-cases: language modeling, image classification, and user behavior modeling implemented using Algorithm~\ref{alg:sampleprog} that is permitted by the policy: 

\begin{align*}
& \texttt{{\color{blue}get\_data} . {\color{blue}train\_local}. 
{\color{red}accumulate$^*$}}\;.\\
& (\texttt{{\color{blue}train\_local}\;.\; {\color{red}accumulate$^*$} + {\color{red} average$^*$})\;.\;return}
\end{align*}

We provide details on tasks, model architectures, training parameters, and convergence
results in Sec.~\ref{sec:use-cases}, and here we only focus on computational and network overhead from using \system.

\begin{figure}[!t]
    \centering
    \includegraphics[width=\linewidth]{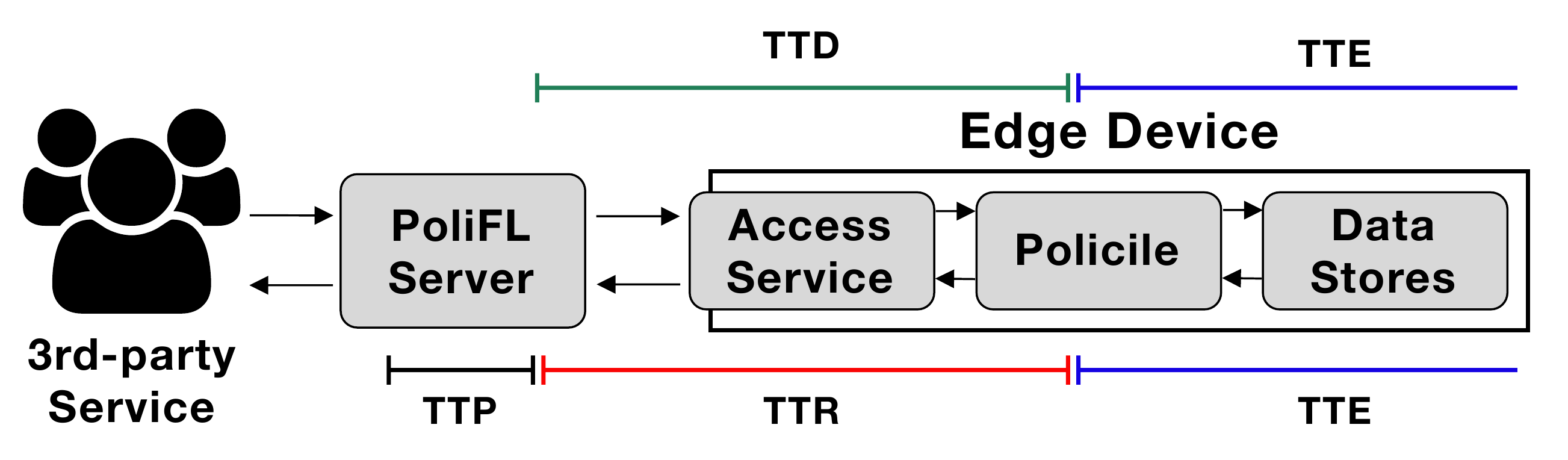}
    \caption{\system Evaluation: We consider measures for the Time to Distribute (TTD), Time to Execute (TTE), Time to Receive (TTR) and Time to Process (TTP).}
    \label{fig:pipeline}
\end{figure}

First, we evaluate the
performance of training on the edge device.  Next, we
measure round trip times for \systemcentral requests executed on an edge
device. In particular, we measure the times shown in
Fig.~\ref{fig:pipeline}: \emph{Time to Distribute} (TTD) is the time
to send the policy program and Data-Policy Pair to the edge device;
\emph{Time to Execute} (TTE) is the time to execute the policy program
on the edge device; \emph{Time to Receive} (TTR) is the time to transmit
the updated Data-Policy Pair back to the \systemcentral; and \emph{Time
to Process} (TPP) is the time to process the received Data-Policy Pair
on the \systemcentral. 

Finally, we evaluate the scalability of the
system by performing a round of training on 100 simulated edge devices measuring TTD, TTR, and TTP on the devices.

\begin{algorithm}[t] \caption{Sample \system training algorithm. \\\hspace{\textwidth}users - list of participants 
\\\hspace{\textwidth}client - remote execution library} 
    \label{alg:sampleprog}
\begin{algorithmic}[1] \Function{remote\_prog}{dpp, model} \State $data$
    = \textbf{get\_data($dpp,~model$)} \State $model$ =
    \textbf{train\_local($model$, $data$)} \State return $model$
    \EndFunction
    \vspace{0.2cm}
    \State $model = $ \textbf{create\_model}()
    \For {round $r$ $\in$ $rounds$} 
    \State $users$
    = \texttt{sample}($total$, m) \For {(user $u$)
    $\gets$ $users$} \State \# create new DPP with user's policy \State
    $dpp = \texttt{\textbf{get\_dpp}}(u) $  \State \# distribute the request
    \State $client.send(model,~dpp,~\texttt{REMOTE\_PROG})$ \EndFor \State $tmp\_sum
    =client.process\_responses()$ \State $model$ =
    \textbf{average}($model$,~$tmp\_sum$) \EndFor \State return $model$
\end{algorithmic}
\end{algorithm}

\subsection{\system Configuration}
\label{sec:poliflconfig}

Our evaluation experiments were run with several hardware and software
configurations described below.  The \systemcentral was installed using Ubuntu Server 18.04 on a Intel i5-8500 6-core processor running at 3.0 GHz with 16 GB of memory and a 256 GB SSD for storage.  

The edge-device overhead evaluations were carried out using a Raspberry~Pi~4 with the specifications listed in Table~\ref{tab:specification} and connected to our \systemcentral on a 1 Gigabit Ethernet Local Area Network (LAN).  As described previously, Policile was used for use-based privacy enforcement and \databox~\cite{10.7146/aahcc.v1i1.21312} as a secure container.

For our scaling tests, we simulate 100 Raspberry~Pi devices connected to the \systemcentral on 1 Gigabit LAN.  We use 50 Intel Core i5-8500 6-core processor machines running at 3.0 GHz with 16 GB of memory. Each machine runs two virtual machines running the same software configuration as on the Raspberry Pi, simulating a total of 100 edge devices.

\begin{table}[!t]
\centering
\caption{Device Specification}
\label{tab:specification}
\begin{tabular}{ ll }

 Type & Specification \\
  \midrule
 Model     & Raspberry Pi 4B \\
 RAM       & 8 GB LPDDR4-3200 SDRAM \\
 Processor & Broadcom BCM2711, Quad-core 1.5GHz \\
 OS        & Raspberry~Pi~OS Lite 64bit (Aug. 2020) \\
 Linux     & Kernel version 5.4 \\
\end{tabular}
\end{table}


\subsection{Edge-Device Overheads}
\label{sec:edgeoverhead}

\begin{figure*}[ht]
    \centering
    \begin{subfigure}[b]{0.32\textwidth}
        \centering
        \includegraphics[width=\textwidth]{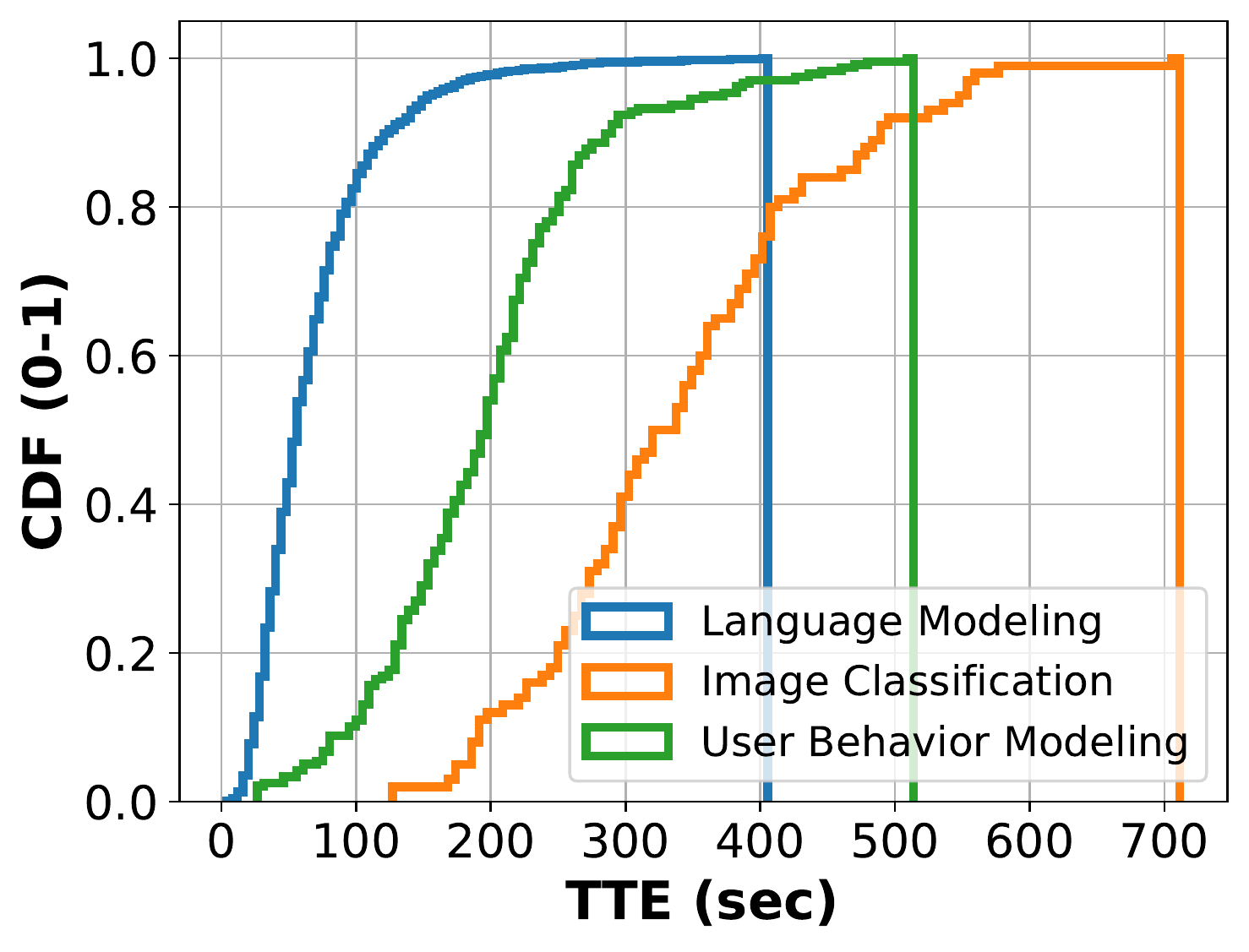}
        \label{fig:cdf_duration_model_uc2}
    \end{subfigure}
    \hfill
    \begin{subfigure}[b]{0.32\textwidth}
        \centering
        \includegraphics[width=\textwidth]{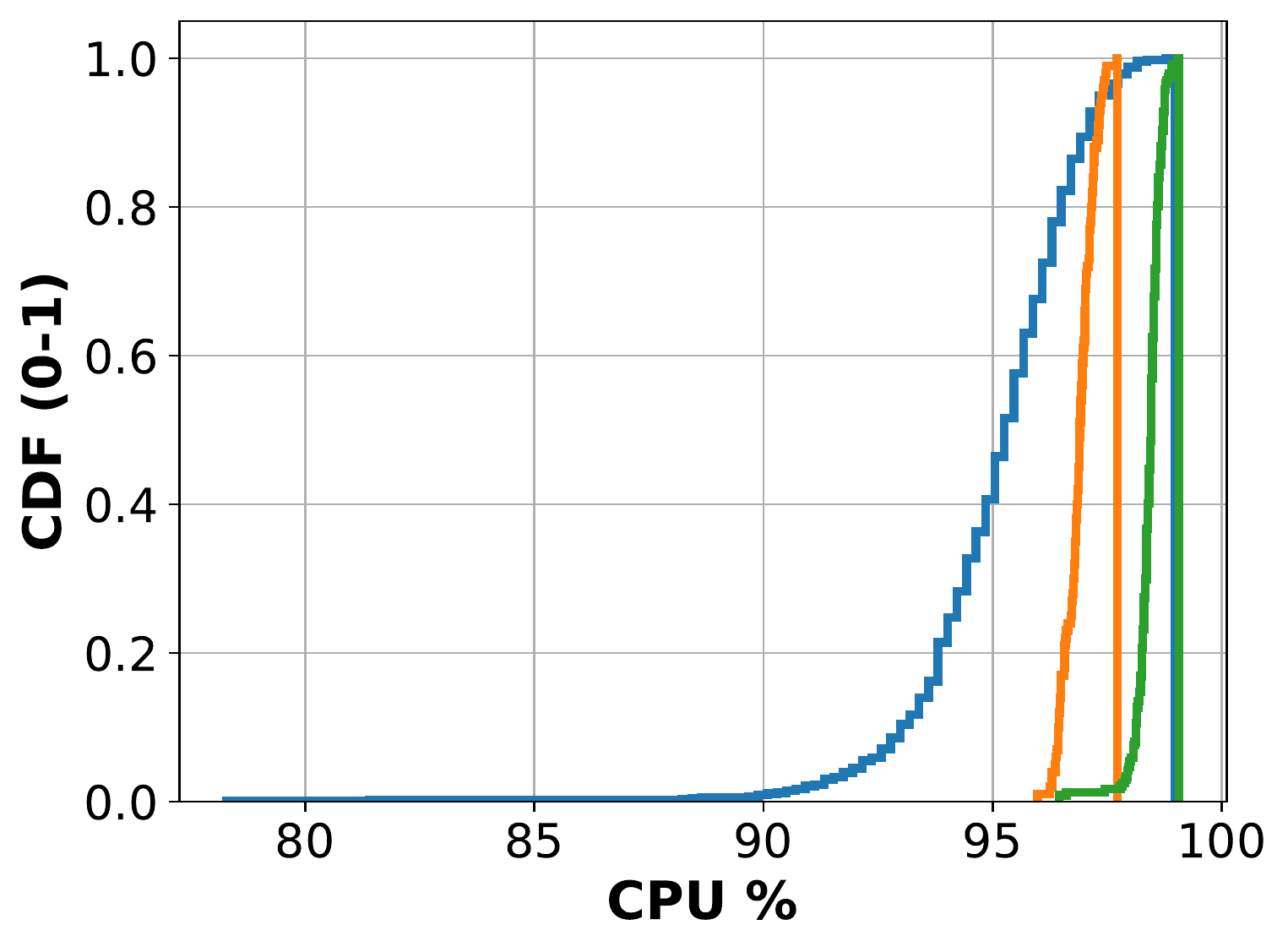}
        \label{fig:cdf_cpu_model_uc2}
    \end{subfigure}
    \hfill
    \begin{subfigure}[b]{0.32\textwidth}
        \centering
        \includegraphics[width=\textwidth]{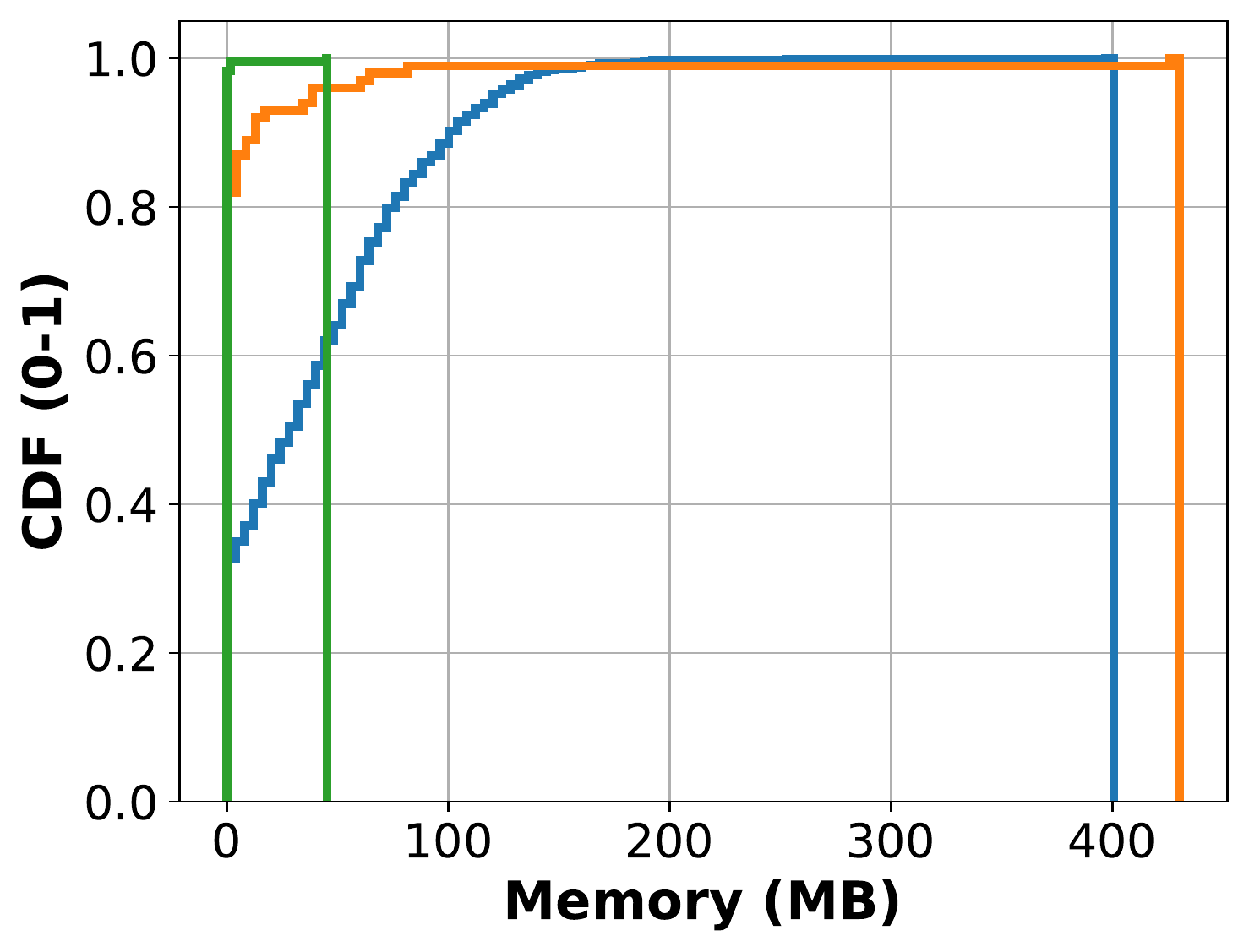}
        \label{fig:cdf_memory_model_uc2}
    \end{subfigure}
    \caption{Cumulative Distribution Function (CDF) representation of the performance metrics (Time to Execute (TTE), CPU utilization and memory overhead) for executing a Policile program (\ie,~training an ML model) for the three use-cases (\ie,~language modeling, image classification, and user behavior modeling) on a single Raspberry Pi edge device.}
    \label{fig:cpu_memory_all}
\end{figure*}

\begin{figure*}[ht]
    \centering
    \includegraphics[width=\textwidth]{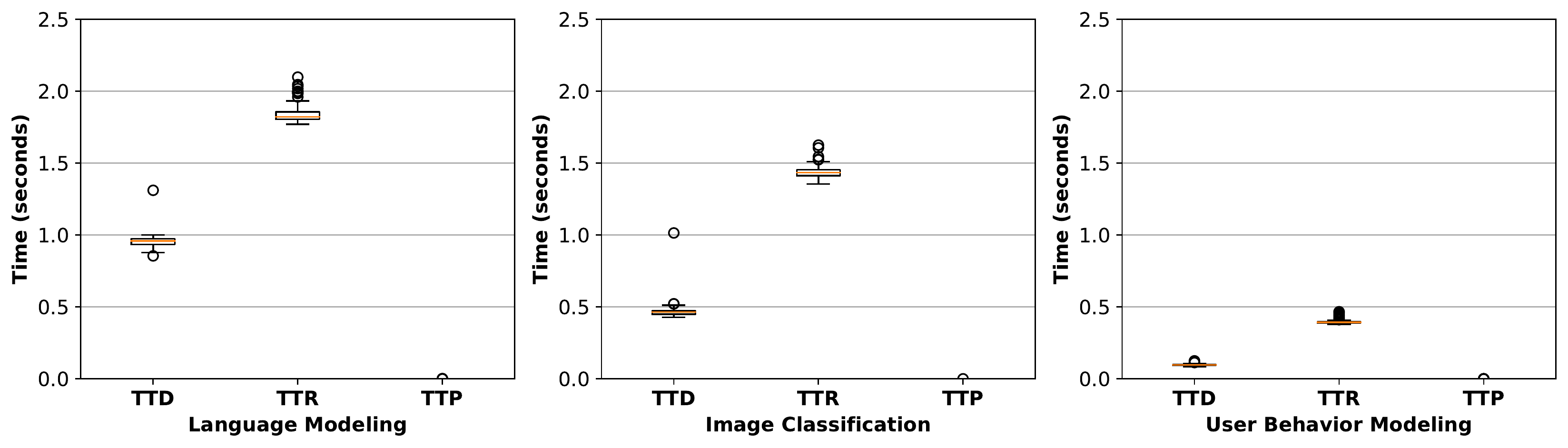}
    \caption{Time to Distribute (TTD), Time to Receive (TTR) and Time to Process (TTP) for our use cases. The TTD and TTR are proportional to the model size and the TTP is negligible.}
    \label{fig:end_to_end}
\end{figure*}

This section reports the performance of an edge device executing a Policile Local Program for training a federating learning model, as previously discussed in
Sec.~\ref{sec:policy-federated-learning}. Specifically, we report
\emph{Time to Execute (TTE)}, the time the model training operation takes to
execute per user, together with the CPU utilization and memory overhead of the
edge device. We run the three referenced tasks with 1000 users for the language modeling task and with all users for the image classification and user behavior modeling tasks (100 and 237, respectively) on a Raspberry~Pi, showing how an edge device with limited resources and with different user data would execute these operations.


Fig.~\ref{fig:cpu_memory_all} summarizes the edge device performance
for this task. Training the model for the language task reported a mean TEE of
$69.62s$~($\pm47.78$), CPU utilization $95.20\%$~($\pm1.75$) and memory
overhead $41.33MB$~($\pm44.29$). The image classification task reported a mean TTE of
$341.50s$~($\pm110.69$), CPU utilization $96.90\%$~($\pm0.32$) and memory
overhead $9.31MB$~($\pm44.78$), while for the user behavior modeling task, a TTE of $197.63s$~($\pm85.36$), CPU utilization of $98.42\%$~($\pm0.31$) and a negligible memory overhead of $0.22MB$~($\pm2.95$). The results show that a limited resources device such as a Raspberry~Pi could be used as an edge device in our
system, even with demanding tasks such as training different types of  models (\ie,~convolutional or LSTM-based networks), with
reasonable execution times and resource overheads.

Note that while the processor was used in almost full capacity during training, the CPU temperature was always below the 80\textdegree{}C threshold where the Pi board starts to reduce the clock speed (\ie,~$56.86$\textdegree{}C ($\pm3.08$)).

\subsection{Network and System Overheads}

Having evaluated the performance of local execution (TTE) on the edge device, we now look at the performance of \system by evaluating networking and processing overheads. We evaluate the system-wide performance for the same three tasks on a single Raspberry Pi edge device. As the previous experiments measured TTE, these experiments measure TTD, TTR, and TTP.

We evaluated the round-trip performance for 100 rounds of training on an edge device. Fig.~\ref{fig:end_to_end} shows 
the range of times for training our three models.  The time to transmit and receive the model is proportional to the size of the model, with the language modeling task having model size of 79 MB, the image classification task having a 43 MB model size, and 12 MB model size for the user behavior modeling task. Our results show that it takes slightly longer to receive the model at the \systemcentral than to distribute it to the edge device.  This is because the model being sent back to \systemcentral from the edge device is slightly larger than the one received due to PyTorch's implementation of running statistics.

The time to process (TPP) the model on the \systemcentral is minimal.  For our tasks it is the time to apply federated averaging to the received models and took an average 0.33 ms and a maximum of 2.2 ms across all rounds for all tasks.  These results show that time for TTD, TTR, and TTP are all negligible compared the time to execute (TTE) the local program on the edge device.

\subsection{Scaling}

Next, we measure the performance of \system using multiple simulated edge devices by performing one round of training with 100 participants for our three modeling tasks. To simulate 100 edge devices, we used 50 workstations, each with the specifications listed in section~\ref{sec:poliflconfig}, and each workstation running two virtual machine (VM) instances of the same software configuration used on the Raspberry Pi. As these workstations are more powerful than the Raspberry Pi, we simulate the TTE for the devices by using the TTE times previously measured on the Raspberry~Pi (Sec.~\ref{sec:edgeoverhead}).

\begin{figure*}[ht]
    \centering
    \includegraphics[width=0.9\textwidth]{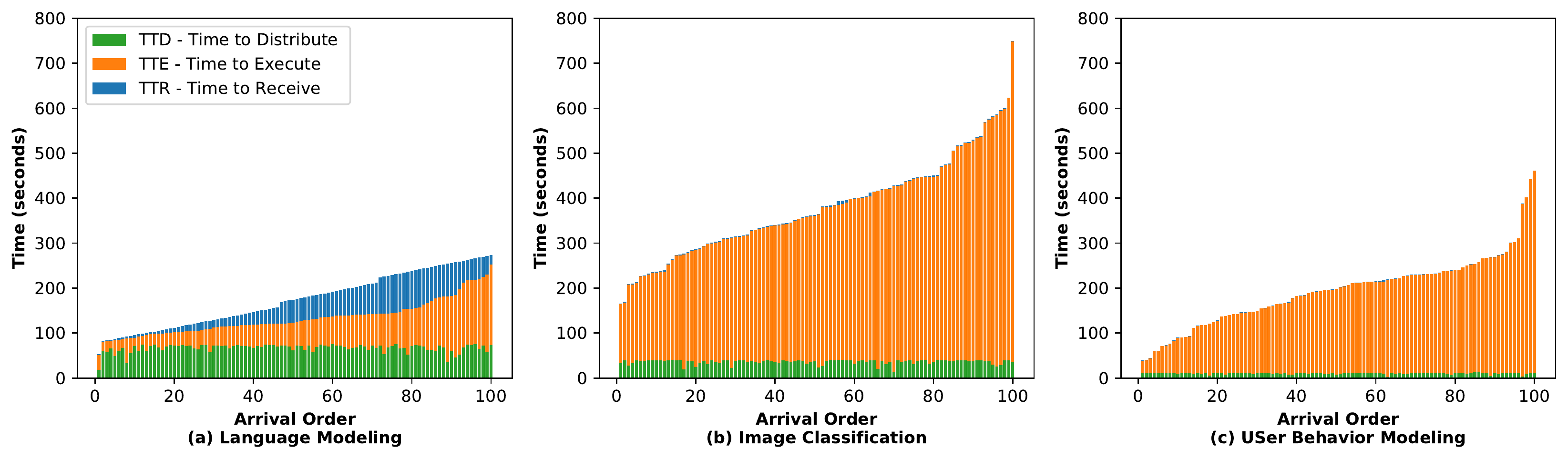}
    \caption{Time to Distribute (TTD), Time to Execute (TTE), and Time to Receive (TTR) for one round of training the ML model for our use cases with 100 simultaneous users. The time to process the request is dominated by the time it takes build the model on the edge devices (TTE). TTD takes longer than TTR because the model is distributed to all the edge devices simultaneously, saturating the network. The Time to Process (TTP) on the \systemcentral is negligible relative to the other times and is not shown here.}
    \label{fig:scaling_100}
\end{figure*}

The \systemcentral distributes the initial model to all 100
devices simultaneously. To match the performance of the edge devices, instead of executing the local program, we sent back a pre-computed model after waiting the amount of time from one TTE times measured in Sec.~\ref{sec:edgeoverhead}.  The results for all three tasks are shown in Fig~\ref{fig:scaling_100}.

The total time to process each device request is shown in the order of arrival at the \systemcentral.  The total time for each request is shown, broken down as TTD, TTE, and TTR.  TTP is not shown as it is negligible compared to the other times.

The results show that it takes longer to distribute the model (TTD) as it does to receive the model (TTR).  This is because the initial model is going out to all 100 devices simultaneously, saturating the network, whereas the computed models are coming back at staggered times.

The majority the time to process the request is due to the time to train the model (TTE), and the limiting factor is the longest time to build the model on some devices. Even in case of the language modeling task, which has the largest model size (79 MB) and shortest average TTE time (69.62s), the overall time to complete the task is gated by the device with the longest model build time.

As the total time to
complete a round of training is dependent on the slowest policy program execution time, the overall time to complete a round of training is not much longer than the TTE of the slowest device. This shows that the longer the local computation on the device, the less important the
networking overheads become at scale.

These results indicate that the main overhead for running Policy-based Federated Learning tasks is firstly the execution time when training an ML model at the edge device, and secondly, with the current implementation of \systemcentral, the bandwidth required for distributing models to multiple edge devices. More advanced distribution techniques could be used in the future to improve the scalability of the system.

\section{Policy-Compliant Federated Learning: Example Use Cases}
\label{sec:use-cases}

In this section, we investigate the performance of our approach when conducting federated learning while following heterogeneous policies. We test three common FL use cases: \emph{language modeling}, \emph{image classification}, and \emph{user behavior modeling}. 
For each use case, we assume that users have one of two different privacy policies drawn from the example policies described in Sec.~\ref{sec:policy-federated-learning:ubp_fl}. 


 \begin{table*}[t]
    \centering
    \caption{\textbf{Policies for different use cases.} Blue color represents execution on the device, red color - execution on the FL platform.}
    \label{tab:policies}
    \begin{tabular}{lrl}
   Use case & Group & Policy \\
    \midrule
    Language Modeling & $Gr_1$ & \texttt{{\color{blue}get\_data(data\_type="reddit")}.{\color{red}runFL.enforce\_privacy\_budget(max\_eps=1)}.return} \\
     & $Gr_2$ & \texttt{{\color{blue}get\_data(data\_type="reddit")}.{\color{red}runFL.enforce\_privacy\_budget(max\_eps=2)}.return} \vspace{0.2cm}\\ 
   Image Classification & $Gr_1$ & 
   \texttt{{\color{blue}get\_data(data\_type="cifar")}.{\color{red}runFL.check\_privacy\_budget(max\_eps=2)}.return} \\
   & $Gr_2$ & 
   \texttt{{\color{blue}get\_data(data\_type="cifar")}.{\color{red}runFL.check\_privacy\_budget(max\_eps=5)}.return} \vspace{0.2cm}\\
   User Behavior Modeling & $Gr_1$  & 
   \texttt{{\color{blue}get\_data(data\_type="MPU")}.{\color{red}runFL}.return} \\
   & $Gr_2$  & 
   \texttt{{\color{blue}get\_data(data\_type="MPU").filter(sensors=[`mic', `loc'])}.{\color{red}runFL}.return} \\
    \end{tabular}
\end{table*}

\paragraphbe{Privacy-Preserving Methods.} In our setting we focus on two main privacy techniques: limiting amount of data
that users share and adding differential privacy to the resulting model. 
We split users into two groups with different privacy regimes and allow third
parties to submit federated learning programs. The third party
is required to satisfy user policies but also needs to maintain acceptable level
of final model performance (since it is trivial to generate a random model that is fully 
private but also unusable).

The main consequence when using differential privacy is that it requires restriction on training
parameters in order to achieve desired budget. In particular, DP benefits from a larger round size and smaller number of total rounds~\cite{fedlearn_dp}. Whereas the round
size is limited by the available memory to accumulate updates from all the participants, the number of total rounds can be reduced at a cost of impacting the model performance. 
Furthermore, if the training involves only part of the dataset, all users participating
in training incur higher privacy costs (see Table~\ref{tab:cascading} for examples on privacy budgets and its dependency on training hyperparameters and dataset size).

\begin{table*}[!ht]
    \centering
    \caption{\textbf{Effect of cascaded training. } Splitting the dataset allows to fully exhaust privacy budgets of each group of users. We compute privacy guarantees using $\delta=10^{{-}8}$ for federated learning with round size = $5000$ that performs short (1000 rounds) or long (2000 rounds) training.}
    \label{tab:cascading}
    \begin{tabular}{rlrrrrrrrr}
     No & Dataset & Budget ($\epsilon_{max}$) & users, $10^8$ & noise & rounds & Privacy spent ($\epsilon$) & Satisfies  \\
    &&&&&&on both groups& budget \\
     \midrule
    1.&  Group $Gr_1$ & $ 1 $      &$1 $       & $0.01$ & $1000$   & $\epsilon^{Gr_1}{=}0.82$, $\epsilon^{Gr_2}{=}0.00$ & Yes \\
    2.&  Group $Gr_2$ & $ 2 $     &$1 $       & $0.005$& $1000$   & $\epsilon^{Gr_1}{=}0.00$, $\epsilon^{Gr_2}{=}1.72$  & Yes \\
    3.&  Group $Gr_1$ (long run) & $ 1 $      &$1 $       & $0.01$ & $2000$   & $\epsilon^{Gr_1}{=}1.08$, $\epsilon^{Gr_2}{=}0.00$ & No \\
    4.&  Group $Gr_2$ (long run) &  $ 2 $    &$1 $       & $0.005$& $2000$   & $\epsilon^{Gr_1}{=}0.00$, $\epsilon^{Gr_2}{=}2.28$ & No \\
    5.&  Combined (long run) & $ min(\epsilon_{max}^{Gr_1}, \epsilon_{max}^{Gr_2}) {=} 1$    &$2 $       & $0.01$ & $2000$   & $\epsilon^{Gr_1}{=}0.82$, $\epsilon^{Gr_2}{=}0.82$ & Yes \\
    6.&  Cascaded (long run) &  $\epsilon_{max}^{Gr_1}{=}1$, $\epsilon_{max}^{Gr_2} {=}2$   &$2 $       & $0.01$ & $2000$   & $\epsilon^{Gr_1}{=}0.82$, $\epsilon^{Gr_2}{=}1.72$ & Yes \\
    \end{tabular}
\end{table*}

\paragraphbe{Policy Controls.} Each use case mentioned above is covered by a policy that
enforces a certain type of training without restricting the model
architecture or training parameters. Table~\ref{tab:policies} shows simple policies, where we use the high-level call \texttt{runFL()} to represent policies corresponding to federated training which can be unraveled to a more detailed view as in Sec.~\ref{sec:policy-federated-learning:ubp_fl}. The \systemcentral requires certain
functions defined in collaboration with the third-party service, \eg,~training, data fetching, and aggregation.  We support
scoping and thus combine multiple training methods under a single
umbrella call \texttt{runFL()}.  We further introduce a function
\texttt{get\_data(data\_type)} that fetches corresponding data
from the participating device.




As stated above, the third party can create their own model and
distribute it to the participants along with the code that should
satisfy the associated policies. The \systemcentral will fetch the
corresponding policies and execute that program. As we stated above,
policies are public, and the provider can design programs that satisfy
them.

\paragraphbe{Policy-Compliance Strategies.} Assuming there exists an ordering on privacy policies (\ie,~more privacy vs. less privacy), there are two natural approaches to training a model in a manner that complies with the (nonuniform) privacy policy associated with each user's data: (1) model training could use training data only from the subset of users with a less-restrictive privacy policy while complying with the less-restrictive privacy policy for that data or (2) model training could use training data from all users while complying with the more-restrictive privacy policy for all users. However, both of these approaches could potentially harm the accuracy of the resulting model: the first strategy both reduces the size of the available training data and introduces the possibility that the training data might be drawn from a different population, while the second strategy is unable to fully utilize all training data to the maximum extent authorized by the privacy policies. 

In an effort to leverage the available training data to the maximum extent authorized by the relevant privacy policies, we introduce a third approach to compliance with policy-based federated learning that we call \emph{cascaded training}. In cascaded training, we first train the model on the subset of users with a more restrictive privacy policy and then fine-tune the resulting model on data from the group with the less-restrictive privacy policy.

We briefly observe that differential privacy is robust to post-processing, which implies that applying fine-tuning on an existing model with new user data (as done in cascaded training) will not impact the privacy budget of the existing users. Cascaded training is therefore compliant with policies that require different levels of differential privacy as long as groups with lower $\epsilon$ values (\ie,~more restrictive privacy policies) are trained separately from groups with higher $\epsilon$ values (\ie,~less restrictive privacy policies).

\paragraphbe{Evaluation Setup.} We evaluate the feasibility of heterogeneous policies for
the three previously described tasks by running full training on the datasets. 
Due to the computational limitations of the Raspberry~Pi deployment when running over
all participants, we resort to 
a generic setup using a server with 4 Nvidia TitanX GPUs. As code and library implementation is the same, we expect the same convergence results to hold when run on a swarm of edge devices. 

We evaluate whether the use of cascaded training is capable of improving 
the performance of the model while preserving heterogeneous privacy restrictions. 
For each use case, we divide the dataset of user data into groups 
with distinct policies from Table~\ref{tab:policies}.

Due to hardware limitations, we cannot 
maintain tight differential privacy guarantees and produce models that converge on an even relatively large dataset 
of $80,000$ Reddit users. Achieving meaningful privacy guarantees on datasets and acceptable utility
require a large number of participants, \eg,~millions of users and large round size shown in 
Table~\ref{tab:cascading} and has been empirically demonstrated by McMahan~\etal~\cite{fedlearn_dp}. 
To mitigate this problem we adopt a notion of weak differential privacy~\cite{naseri2020toward}, aiming to mimic the DP parameters, \eg,~clipping bound and noise distribution, proposed by McMahan~\etal~\cite{fedlearn_dp}.

\subsection{Language modeling}

Typed text can include sensitive information. Distributed training can help conceal the data by releasing only trained models instead of raw text~\cite{hard2018federated}. 

We use our system to train a word prediction model using the Reddit dataset~\cite{fedlearn_1}. We use $80,000$ posts from the public Reddit dataset considering users that have between 150 and 500 posts with 247 posts each on average. The task is to predict the next word given a partial word sequence. Each post is treated as a training sentence. We use a two-layer, 10M-parameter sample LSTM model~\cite{pytorchwordmodel} with 200 hidden units. An input is split into a sequence of 64 words. For participants local training, we use a batch size of 20, learning rate of 20, and the SGD optimizer. 

We run up to 6,000 rounds of training and picking 100 participants every round. A model 
trained on all user data without any privacy restrictions achieves total accuracy of $19\%$.
For differential private training we follow parameters from~\cite{fedlearn_dp} and set the clipping bound $S=15$. For one experiment, users were divided into two equal groups (\emph{$Gr_1$} and \emph{$Gr_2$}) with added noise $\sigma$ of 0.01 and 0.001, respectively.  For the second experiment,  users were divided into three equal groups (\emph{$Gr_1$}, \emph{$Gr_2$}, and \emph{$Gr_3$}) with added noise $\sigma$ of 0.01, 0.005, and 0.001, respectively. As differential privacy is sensitive to a number of total rounds
we restrict training to a maximum allowed rounds for the whole dataset.

\begin{figure*}[!ht]
\captionsetup[subfigure]{justification=centering}
  \centering
  \begin{subfigure}{0.9\textwidth}
    \centering
    \includegraphics[width=\linewidth]{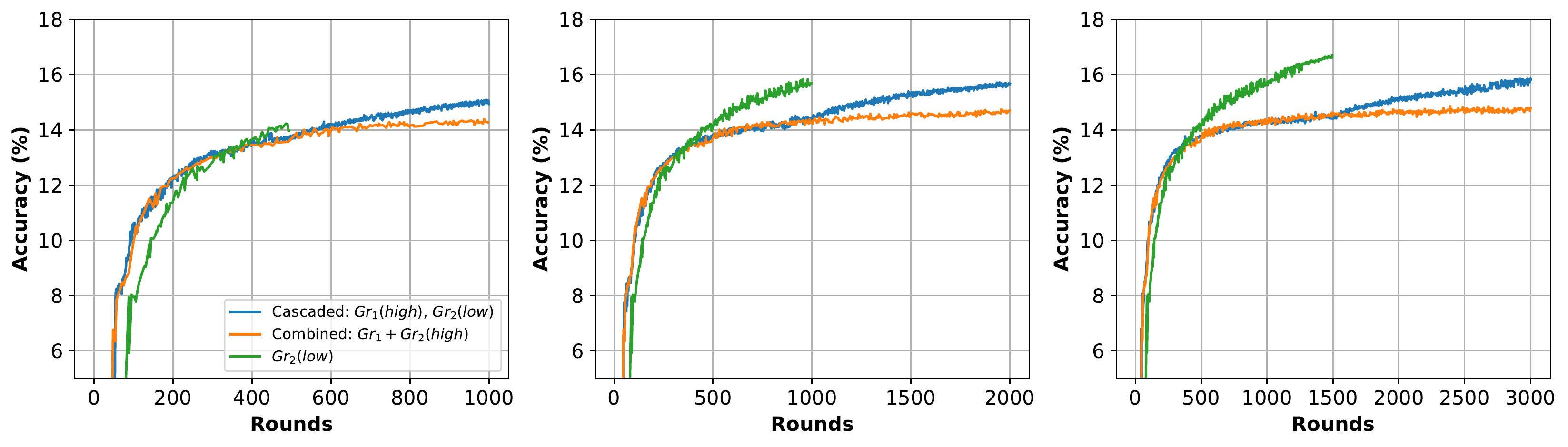}
    \caption{Two privacy groups}\label{fig:cascade:two}
  \end{subfigure}
  \newline
  \begin{subfigure}{0.9\textwidth}
      \centering
      \includegraphics[width=\linewidth]{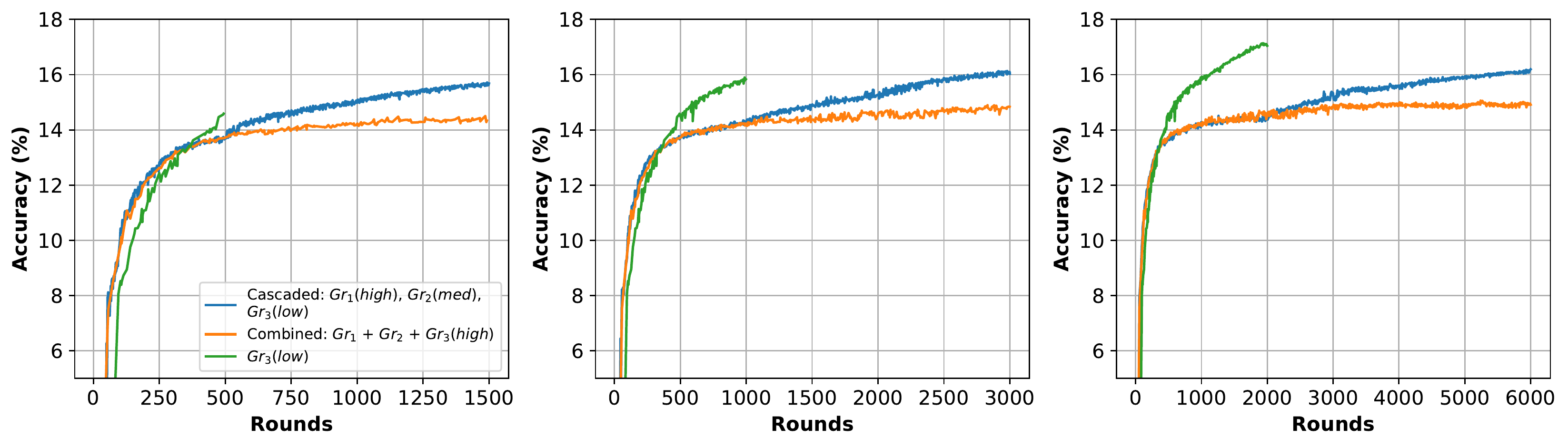}
      \caption{Three privacy groups}\label{fig:cascade:three}
  \end{subfigure}
  \caption{\textbf{Performance of the language modeling task.} Evaluating performance on users split into (a) two privacy groups, and (b) three privacy groups. The cascaded  experiment trained with the highest privacy group first ($\sigma {=} 0.01,S{=}15$) and the lowest privacy group last ($\sigma {=} 0.001,S{=}15$).  The three group cascaded experiment also had a medium privacy group ($\sigma {=} 0.005,S{=}15$). The combined experiment had all groups trained using high privacy setting ($\sigma{=}0.01,S{=}15$). The last experiment trained the model using data only from the low privacy group ($\sigma {=} 0.001, S{=}15$) but trained for fewer rounds to maintain differential privacy guarantees.}
  \label{fig:cascade:both}
\end{figure*}

\paragraphbe{Results.} The results for the two-group and three-group cascade
are shown in Fig.~\ref{fig:cascade:two} and \ref{fig:cascade:three}, respectively. Our experiments are based on the
settings shown in Table~\ref{tab:cascading}, but using a smaller dataset.
We compare the cascading trained model (row 6) to two other models, one trained with all data but at a higher privacy (lower $\epsilon$ budget, row 5) 
and the other using the low privacy (higher $\epsilon$ budget, row 2) group. We ignore rows 3 and 4 as they do not satisfy privacy budgets, as well as the row 1 case that is part of a combined case 5.
Following the results in Table~\ref{tab:cascading} (rows 5 and 6)
when two groups are combined together (\eg,~using the complete dataset) we run FL for twice as long.

In all cases, training using all data but at the higher
privacy setting (row 5) was the worst performing model.  The cascaded
model (row 6) outperforms the low privacy model on shorter training
experiments
but does not perform well when longer training is allowed.  In the
two groups case (Fig.~\ref{fig:cascade:two}), training the cascaded model for 1000 rounds
resulted in better performance.  When training the cascaded model for
2000 rounds, accuracy was similar to the low privacy model 
trained for 1000 rounds. Training the cascaded model for
3000 rounds gave lower performance than training the low privacy 
models for 1500 rounds.  We see similar results for the three
group experiments (Fig.~\ref{fig:cascade:three}).  The cascaded model does better than the
low privacy model at 1500 rounds, similar accuracy for 3000
rounds of training, and lower accuracy at 6000 rounds.  These
results show that a cascaded approach works well for models
that cannot be trained for many rounds due to their privacy budget restriction, however
when the training is allowed for more rounds training solely on a single group with less
added noise achieves better performance.

\subsection{Image Classification}
A locally trained image classification model is useful when a user might want to keep photos private on their local device but still decide to contribute to the global recognition model.
 
We use CIFAR-10 image recognition~\cite{krizhevsky2009learning} task as another scenario of use for Federated Learning. We split the data among 100 participants using Dirichlet distribution with $\alpha=0.9$. We use ResNet18 model~\cite{he2016deep} with $11$ mln parameters, batch size of 32, and learning rate of $0.1$. We run 350 rounds of training selecting 10 participants at each round and $\eta=5$. For differentially private federated learning we set the clipping bound to $S=15$. Users were divided into two equal groups (\emph{$Gr_1$} and \emph{$Gr_2$}) of high and low privacy with added noise $\sigma$ of 0.01 and 0.001, respectively.

\begin{figure}[t!]
    \centering
    \captionsetup[subfigure]{justification=centering}
    \includegraphics[width=.9\linewidth]{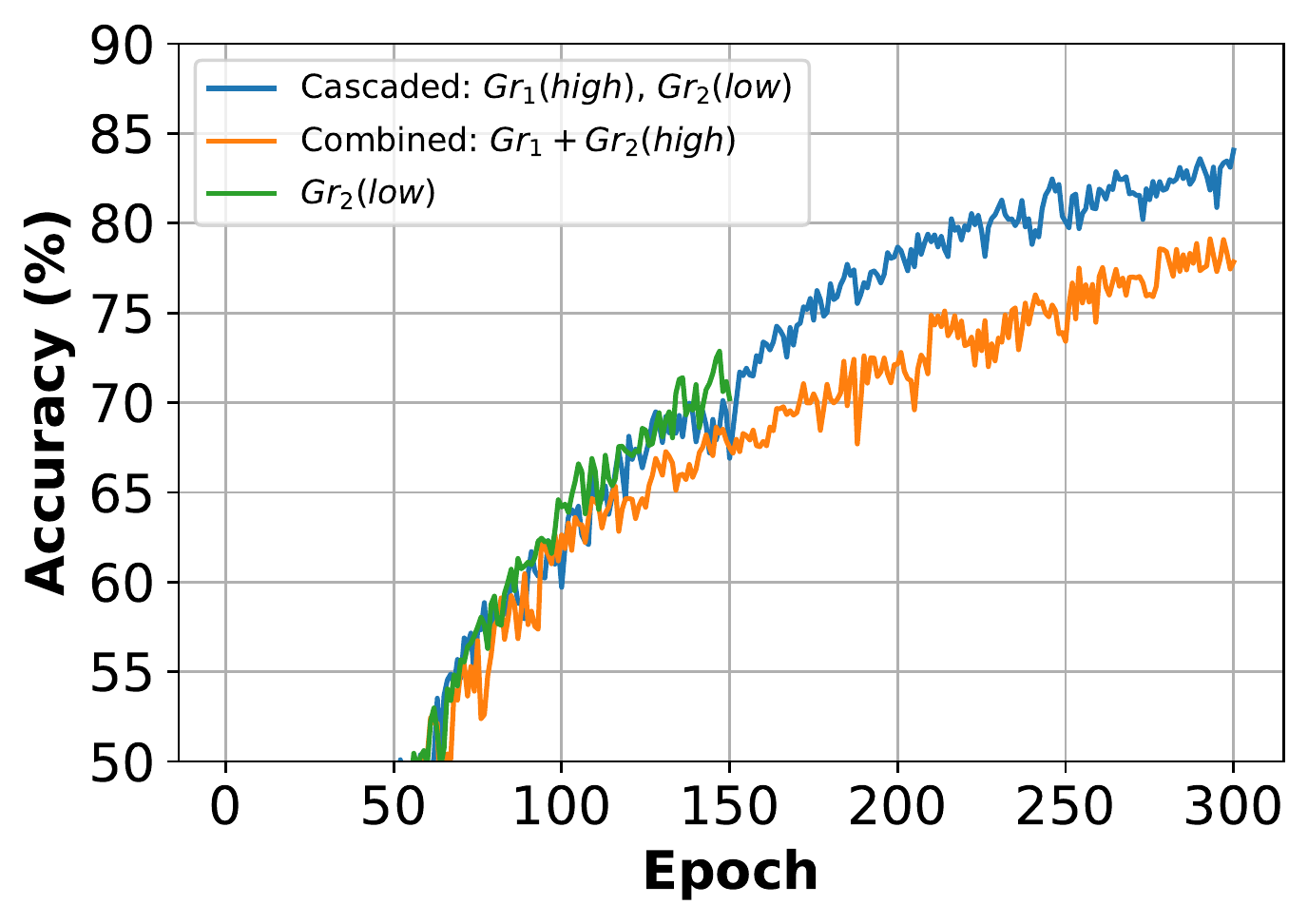}
    \caption{\textbf{Performance of the image classification task.} A model trained with cascaded training method outperforms a model trained on combined data but with high privacy ($\sigma{=}0.01, S{=}15$) and trained on a single group but with less privacy ($\sigma{=}0.001, S{=}15$).}
    \label{fig:cascade_uc2}
\end{figure}

\paragraphbe{Results.} Fig.~\ref{fig:cascade_uc2} presents the training performance of the two privacy groups, together with the case of cascaded training. In contrast to the previous use case (language modeling task), results of this task show that cascaded training outperforms the cases when all users were set with same high or low privacy budget ($\sigma=0.01$ or $\sigma=0.001$, respectively). Training for users with low privacy preference could not exceed epoch 150 in order to keep the group's privacy budget valid, and consequently achieved a performance 70.22\%. Users with high privacy budget achieved a higher accuracy of 77.84\% where cascaded training outperformed both other cases, reached an accuracy of 84.07\%.

\subsection{User Behavior modeling}
This use case predicts whether a user will engage with a notification (\ie,~click at the notification or open the corresponding app) within 10 minutes it is received from their device~\cite{katevas2017practical}.

We used the Mobile Phone Use (MPU) dataset~\cite{pielot2017beyond} which includes use-logs from 237 Android phone users for an average duration of four weeks. More particularly, it includes data from 15 physical and virtual mobile sensors (\ie,~accelerometer, battery, network data, light, noise, location, app used, audio source, music played, charging state, notification, access to notification center, ringer mode, screen state, screen orientation).
 
Following the work from~\cite{katevas2017practical}, we use a fully-connected time-distributed linear layer with 50 parametric rectified linear units, two stateful LSTMs as hidden layers with 500 units, and a dense layer with sigmoid activation function and total of 
$3$ mln parameters. For differentially private federated learning we similarly set $S=15$. Note that for this task, due to the unbalanced nature of the dataset, we first computed the receiver operating characteristics (ROC) curve and report the performance as the area under the curve (AUC), also suggested in the original work~\cite{katevas2017practical}.

In this task we tested several different cases, listed in Table~\ref{tab:tf_perf}. We divided users into two equal groups ($Gr_1$ and $Gr_2$) and manipulated (a) the data access (users either contribute all their data, or exclude location and microphone) and (b) the preference of high and low privacy, with added noise $\sigma$ of 0.01 and 0.001, respectively.
For more information about the executed policies, please refer to Table~\ref{tab:policies}.
 
\begin{table}[t]
    \centering
    \caption{\textbf{Performance (in ROC~AUC) of user behavior modeling task}. We manipulated (a) the sensor data access, and (b) the privacy preference. Training was performed for 500 rounds with 10 participants per round.}
    \label{tab:tf_perf}
    \begin{tabular}{clc}
    & Use case & Perf. \\
   \midrule
   \multirow{4}{*}{\rotatebox[origin=c]{90}{Sensors}}
    & $Gr_1+Gr_2$ (all data) & 72.4\% \\
    & $Gr_1$ (all data), $Gr_2$ (no mic+location) &  70.6\% \\
    & $Gr_1$ (no mic+location), $Gr_2$ (all data) & 71.1\% \\
    & $Gr_1+Gr_2$ (no mic+location) & 69.8\% \\
    \midrule
    \multirow{3}{*}{\rotatebox[origin=c]{90}{DP}}
    & Combined: $Gr_1$ (high privacy), $Gr_2$ (low privacy) & 70.9\% \\
    & Cascaded: $Gr_1+Gr_2$ (high privacy) & 68.4\% \\
    & $Gr_2$ (low privacy, stopped at epoch 250) & 69.1\% \\
    \end{tabular}
\end{table}

\paragraphbe{Results.} Table~\ref{tab:tf_perf} shows the performance of the user behavior modeling task in the two different cases of policy preference. Our approach for cascaded training in the example of data access-based policies was to train the group with less data first (\ie,~no microphone and location data) and further personalize the model with the group of richer data (\ie,~include all data), achieving an AUC performance of $71.1\%$, a $1.13\%$ increase compared to a homogeneous policy that does not allow access to these sensitive data for all users ($Gr_1+Gr_2$). Similar to the other two tasks, cascaded training achieved an accuracy of 70.9\%, outperforming both other cases of high (68.4\%) and low (69.1\%) privacy preference. Note that, once again, the low privacy task had to be stopped at epoch 250 in order to keep the $\epsilon$ guarantee valid.

\section{Conclusions}
\label{sec:conclusions}

We present \system, a decentralized policy-based framework for federated learning tasks. Using a number of
privacy-sensitive use cases including text, image and mobile sensor-based models, we
evaluated \system and demonstrated its feasibility and scalability.


Further work will be required to identify the full range of policies appropriate
for federated learning and to provide formal guarantees for how different
policies interact. Nonetheless, this work constitutes and important step
towards supporting general heterogeneous policies for federated learning.

\begin{acks}
    The authors would like to thank Nate Foster and Fred B.
Schneider for the initial productive discussions
and ideas. This work was supported in part by the NSF Grant 1642120.
Haddadi and Katevas were partially funded by the EPSRC Databox project
EP/N028260/1 and the EPSRC DADA project EP/R03351X/1. 
\end{acks}

\bibliographystyle{abbrv}
\bibliography{references}

\appendix
\clearpage
\section*{Appendix}
\label{sec:Appendix}

\section{Policy Enforcement}
\label{appx:policyenforce}

We adopt the same structure of policy enforcement as
Ancile~\cite{bagdasaryan2019ancile}: we treat a policy as a regular 
expression and advance it using modified version of Brzozowski derivatives~\cite{brzozowski1964derivatives}, see notation on
Fig.~\ref{fig:grammar}.
A summary of the
operations on policies $E$ is given in Fig.~\ref{fig:estep} and the derivative step is given in Fig.~\ref{fig:dstep}.



\begin{figure}[b]
\vspace{-\baselineskip}
		\begin{align*}
		& P &::= \; &C&                   &-\; &\texttt{command} \\
		&    &| \; &P_1 \; . \; P_2&     &-\; &\texttt{sequence} \\
		&    &| \; &( P_1 \; + \; P_2)&  &-\; &\texttt{union} \\
		&    &| \; &( P_1 \; \& \; P_2)& &-\; &\texttt{intersection} \\
		&    &| \; &!P&                  &-\; &\texttt{negation} \\
		&    &| \; &P^*&                 &-\; &\texttt{Kleene star} \\
		&    &|\; &0&                   &-\; &\texttt{no operation} 		
		\end{align*}
	\caption{Policy Syntax}
	\label{fig:grammar}

        \begin{align*}
        & E(\texttt{0}) &=&\;\; 0 \\
        & E(C) &=&\:\: 0 \\
        & E(P_1 \; . \;  P_2) &=&\;\; E(P_1) \; \wedge \; E(P_2) \\
        & E(P_1 \; + \;  P_2) &=&\;\; E(P_1) \; \vee \; E(P_2) \\
        & E(P_1 \; \& \; P_2) &=&\;\; E(P_1) \; \wedge \;  E(P_2) \\
        & E(P^*)              &=&\;\; 1 \\ 
        & E(!P)               &=&\;\; !E(P) \\
        & & & 
        \end{align*}
    \caption{Emptiness check operation $E(P)$}
    \label{fig:estep}

		\begin{align*}
		& D(\texttt{0},C)       &=&\;\;   \texttt{0} \\
		& D(C,C)                &=&\;\;   \texttt{1} \\
		& D(C, C')              &=&\;\;   \texttt{0} \textbf{ (for }C \neq C') \\
		& D(P_1 \; . \; P_2, C) &=&\;\;   D(P_1,C) . P_2  {+} E(P_1) \; . \; D(P_2, C) \\
		& D(P_1 {+} P_2, C) &=&\;\;   D(P_1, C) \; + \; D(P_2, C) \\
		& D(P1 {\&} P2, C)  &=&\;\;   D(P_1,C) \; \& \; D(P_2, C) \\
		& D(P^*, C)             &=&\;\;   D(P,C) \; . \; P^* \\
		& D(!P, C)              &=&\;\;   !D(P,C)
		\end{align*}
	\caption{A summary of the syntactic operation D}
	\label{fig:dstep}
\end{figure}

 
 


\begin{figure}[b]
\vspace{-\baselineskip}

\begin{minipage}{\columnwidth}
 \begin{algorithm}[H]
    \caption{Federated Learning: Local training.} 
    \label{alg:trainlocal}
    \begin{algorithmic}[1]
        \Function{train\_local}{model $G^t$, data $\mathcal{D}_{local}$} 
        \State \textit{\# Initialize local model $L$ for time $t{+}1$} 
        \State $L^{t+1} \leftarrow G^t $ 
        \For{epoch $e\gets 1, E$}
        \For{batch $b$  $\in \mathcal{D}_{local}$} 
            \State  $L^{t+1} \leftarrow L^{t+1} - lr \cdot
        \nabla  \ell  (L^{t+1} , b)$ 
        \EndFor 
        \EndFor 
        \State \textbf{return} $L^{t+1}$
        \EndFunction
     \end{algorithmic} 
\end{algorithm}

\end{minipage}

\begin{minipage}{\columnwidth}
\vspace*{-\baselineskip}
 \begin{algorithm}[H]
    \caption{Addition of Differential Privacy.} 
    \label{alg:traindp}
 \begin{algorithmic}[1]
    \Function{train\_local\_dp}{$G^t$, $\mathcal{D}_{local}$}: 
    \State \textit{\# Perform normal local training}
    \State $L^{t+1} \leftarrow \texttt{train\_local}(model, \mathcal{D}_{local}) $ 
    
    \State \textbf{return} \texttt{Clip($L^{t+1}, S)$} + $\mathcal{N}(0, \sigma^2)/m$
    \EndFunction
 \end{algorithmic} 
 \end{algorithm}
\end{minipage}

\begin{minipage}{\columnwidth}
\vspace*{-\baselineskip}
 \begin{algorithm}[H]
    \caption{Federated Learning: Accumulate local models.} 
    \label{alg:accumulate}
 \begin{algorithmic}[1]
    \Function{accumulate}{model $L$, $tmp\_sum$}
    \For {name $n$, param $p$ $\in$ $L.parameters()$} 
        \State $tmp\_sum[n] {+}{=} p$ 
    \EndFor
    \State \textbf{return} $tmp\_sum$
    \EndFunction
 \end{algorithmic} 
 \end{algorithm}
\end{minipage}

\begin{minipage}{\columnwidth}
\vspace*{-\baselineskip}
\begin{algorithm}[H]
 \begin{algorithmic}[1]
    \caption{Federated Learning: Updating global model.} 
    \label{alg:average}
    \Function{average}{global model $G^t$, $sum^{t+1}$} 
    \For {name $name$, param $p$ $\in$ $G^t.parameters()$} 
        \State $update = \eta/n \cdot (sum^{t+1}[name])$
        \State $param.add\_(update)$ 
    \EndFor 
    \State $G^{t+1} = G^t$
    \State return $G^{t+1}$
    \EndFunction
 \end{algorithmic} 
\end{algorithm}
\end{minipage}
\vspace{-0.1cm}
\end{figure}

\end{document}